\documentclass[twocolumn,pra,showpacs,superscriptaddress]{revtex4}
\usepackage{amssymb}
\usepackage{amsmath}
\usepackage{graphicx}
\usepackage{subfigure}
\usepackage{natbib}
\usepackage{epsfig}
\usepackage{amsfonts}
\usepackage{mathrsfs}
\usepackage{CJK}
\usepackage[toc,page,title,titletoc,header]{appendix}

\begin{document}

\title{Scattering and bound states of two polaritons in an array of coupled
cavities}
\author{Chuan-zhou Zhu}
\affiliation{Department of Physics, Renmin University of China, Beijing, 100872, China}
\author{Shimpei Endo}
\affiliation{Department of Physics, University of Tokyo, 7-3-1 Hongo, Bunkyo-ku, Tokyo
113-0033, Japan}
\author{Pascal Naidon}
\affiliation{RIKEN Nishina Centre, RIKEN, Wako 351-0198, Japan,}
\author{Peng Zhang }
\email{pengzhang@ruc.edu.cn}
\affiliation{Department of Physics, Renmin University of China, Beijing, 100872, China}

\begin{abstract}
We develop an analytical approach for calculating the scattering and bound
states of two polaritons in a one-dimensional (1D) infinite array of coupled
cavities, with each cavity coupled to a two-level system (TLS). In particular, we
find that in such a system a contact interaction
between two polaritons is induced by the nonlinearity of the
Jaynes-Cummigs Hamiltonian.
Using our
approach we solve the two-polariton problem with zero center-of-mass momentum, and find 1D resonances. Our results are relevant to the transport of two polaritons, and are helpful for
the investigation of many-body physics in a dilute gas of polaritons in a 1D
cavity array.
\end{abstract}

\maketitle

\section{Introduction}

In the recent years, the investigation of the physics in one-dimensional
(1D)~\cite%
{1d1,1d2,1d3,1d4,1d5,1d6,1d7,1d8,1d9,1d10,1d11,1d12,1d13,1d2d1,1d2d2} and
two-dimensional (2D)~\cite{1d2d1,1d2d2,2d1,2d2,2d3,2d4,2d5,2d6,2d7,2d8}
array of coupled cavities has attracted a lot of attention. It is predicted
that, such systems can be used in quantum information processing~\cite%
{1d1,1d2,1d3,1d4,1d5} as well as the quantum simulation of many-body
systems, e.g., the quantum phase simulation~\cite%
{1d2d1,1d2d2,1d10,1d11,1d13,2d1,2d2,2d3,2d4,2d5,1d12}, quantum Hall effect~%
\cite{2d6} and Bose-Einstein condensate~\cite{2d7}. For the few-body physics
of coupled cavities, many authors have studied the single-photon
transmission in a 1D cavity array coupled with a single atom~\cite{1d1}, and
the dynamics of a single polariton in a 1D cavity array with each cavity
coupled to an atom~\cite{1d2,1d7}. Recently, bound states of two polaritons
in such a system with finite number of cavities was studied by Wong and Law
by direct numerical diagonalization of the Hamiltonian~\cite{1d8}.
However, the scattering and bound states of two polaritons in an infinite
array of coupled cavities remain to be investigated.

In this paper, we develop an analytical approach for calculating the
scattering and bound states of two polaritons in an infinite array of
coupled cavities with each cavity coupled to a two-level system (TLS), which
can be either a natural or an artificial
atom. In particular, we find that in such a system there is an effective contact interaction between
two polaritons. Similarly to the photon blockade phenomenon~\cite{pb}, this
effective interaction is also due to the nonlinearity of the
Jaynes-Cummigs (JC) Hamiltonian.
With our approach, one can treat the problem with standard techniques of
quantum scattering. We further derive the scattering coefficient and
bound-state energy of two polaritons with zero center-of-mass momentum. The
resonance phenomena induced by the weakly bound states are also
investigated. Our results can be directly used in the research of the
transport of two polaritons. It is also helpful for the investigation of the
many-body physics in a dilute gas of polaritons in a 1D cavity array.

This paper is organized as follows. In Sec. II we derive the effective
interaction between two polaritons. Based on this result, the analytical
method for calculating two-polariton scattering states and bound states is
developed in Sec. III. In Sec. IV we solve the two-polariton problem with
zero center-of-mass momentum with our approach and analyze the resonance
phenomena. We conclude and discuss these results in Sec. V.

\section{Two-body problem of polaritons in cavity array}

We consider a one-dimensional (1D) infinite array of coupled single-mode
cavities. In each cavity there is a two-level system (TLS), which interacts
with the photon in the cavity.

\begin{figure}[tbh]
\centering
\includegraphics[bb=35bp 535bp 563bp 749bp, width=8cm, clip]{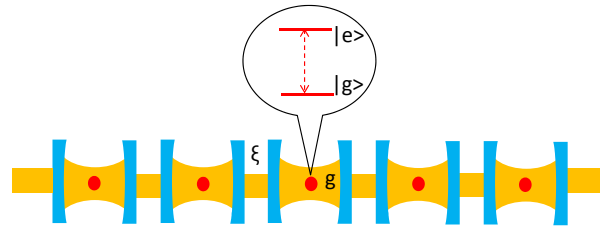}
\caption{(Color online) Schematic illustration of a one-dimensional infinite
array of coupled single-mode cavities, with each cavity coupled to a TLS.
The single-photon hopping intensity between the adjacent cavities is $%
\protect\xi$, and the coupling intensity between the TLS and the photon in
the same cavity is $g$.}
\label{fig1}
\end{figure}

In the interaction picture, the Hamiltonian of the system is given by%
\begin{equation}
H=\Delta \sum_{n=-\infty }^{\infty }\sigma _{ee}^{(n)}+g\sum_{n=-\infty
}^{\infty }\sigma _{eg}^{\left( n\right) }a_{n}+\xi \sum_{n=-\infty
}^{\infty }a_{n+1}^{\dag }a_{n}+h.c.\,,  \label{H}
\end{equation}%
where $\Delta=\omega_A-\omega_L$ is the detuning between the TLS and the
cavity mode, with $\omega_{A}$ and $\omega_{L}$ the frequency of the TLS and
the photon, respectively. In Eq.~(\ref{H}) $\xi $ is the hopping intensity
or the inter-cavity coupling strength, $a_{n}$ and $a_{n}^{\dag }$ are the
annihilation and creation operators of the photons in the $n$-th cavity
respectively, and $g$ is the coupling intensity between the TLS and the
photon in the $n$-th cavity. Without loss of generality, here we assume that
$g$ and $\xi $ are real numbers. The Pauli operator $\sigma _{ij}^{(n)}$ ($%
i,j=e,g$) is defined as $\sigma _{ij}^{(n)}=|i\rangle _{t}^{(n)}\langle j|$,
with $|g\rangle _{t}^{(n)}$ and $|e\rangle _{t}^{(n)}$ the ground and
excited state of the TLS in the $n$-th cavity, respectively. For the
convenience of our following discussion, we further define $|0\rangle
_{c}^{(n)}$ as the vacuum state of the $n$-th cavity.

We can define the \textquotedblleft vacuum" state of our total system as%
\begin{equation}
|G\rangle =\prod\limits_{n=-\infty }^{\infty }|g\rangle _{t}^{(n)}|0\rangle
_{c}^{(n)}
\end{equation}%
where all the atoms are in their ground level and there is no photon. Both
the creation of photon and the excitation of TLS can be considered as the
polaritons of the system. We can define the number operator of polaritons as%
\begin{equation}
N=\sum_{n=-\infty }^{\infty }\sigma _{ee}^{\left( n\right) }+a_{n}^{\dag
}a_{n}.
\end{equation}%
It is clear that $\left[ H,N\right] =0$ and the polariton number is
conserved. In case of $N=1$, there is a single polariton in our system, and
Hamiltonian $H$ has eigen-states $A_{k}^{\dagger }|G\rangle $ or $%
B_{k}^{\dagger }|G\rangle $ with eigen-energies $\varepsilon _{Ak,Bk}=(2\xi
\cos k+\Delta \pm \sqrt{(2\xi \cos k-\Delta )^{2}+4g^{2}})/2.$ Here
\begin{equation}
A_{k}^{\dagger }(B_{k}^{\dagger })=\sum_{n=-\infty }^{\infty }e^{ikn}\left[
\eta _{A(B)}\left( k\right) \sigma _{eg}^{\left( n\right) }+\eta
_{A(B)}^{\prime }\left( k\right) a_{n}^{+}\right]  \label{ab}
\end{equation}%
are the creation operators of the excitonic polariton of kind $A$ or $B$
with momentum $k$. The coefficients $\eta _{A,B}\left( k\right) $ and $\eta
_{A,B}^{\prime }\left( k\right) $ can be obtained straightforwardly from the
eigen-equaiton of $H$.

In this paper we consider the two-body problem of polaritons in our system,
i.e., the quantum dynamics of our system in the subspace with $N=2$. In such
a subspace, due to the translation symmetry of our system, the total
momentum of the two polaritons is conserved, and thus the eigen-state of the
Hamiltonian $H$ can be written as
\begin{eqnarray}
|\Psi \rangle  &=&\sum_{m,n=-\infty }^{\infty }\frac{1}{\sqrt{2}}e^{iK\left(
m+n\right) /2}p_{m-n}a_{m}^{\dag }a_{n}^{\dag }|G\rangle   \notag \\
&&+\sum_{m,n=-\infty }^{\infty }e^{iK\left( m+n\right) /2}d_{m-n}a_{m}^{\dag
}\sigma _{eg}^{(n)}|G\rangle   \notag \\
&&+\sum_{m,n=-\infty }^{\infty }\frac{1}{\sqrt{2}}e^{iK\left( m+n\right)
/2}t_{m-n}\sigma _{eg}^{(m)}\sigma _{eg}^{(n)}|G\rangle ,  \label{psi}
\end{eqnarray}%
with $K\in \lbrack -2\pi ,2\pi )$ the total momentum of the two polaritons,
and the coefficients $t_{l},d_{l},p_{l}$ describe the \textit{relative motion%
} of the two polaritons, and satisfy
\begin{equation}
p_{l}=p_{-l},\ \ \ t_{l}=t_{-l},\ \ \ t_{0}=0\,.\label{six}
\end{equation}%
We can further define the coefficients $d_{l\pm }=\left( d_{l}\pm
d_{-l}\right) /2$ which satisfy $d_{l\pm }=\pm d_{-l\pm }.$

It is apparent that the quantum state $|\Psi \rangle $ is described by a $4$%
-dimensional vector
\begin{equation}
\beta _{l}=(p_{l},d_{l+},d_{l-},t_{l})^{T}\ \ \ \ (l=0,\pm 1,\pm 2,...),
\end{equation}%
which can be considered as the ``spinor wave function" of the relative
motion of two polaritons in the state $|\Psi\rangle$.  Thus, the condition (%
\ref{six}) for the coefficients can be re-expressed as
\begin{equation}
\beta _{l}=\mathcal{T}\beta _{-l};\ \ \ \ \ t_{0}=0.  \label{c1}
\end{equation}%
Here the matrix $\mathcal{T}$ is defined as
\begin{equation}
\mathcal{T=}\left(
\begin{array}{cccc}
1 & 0 & 0 & 0 \\
0 & 1 & 0 & 0 \\
0 & 0 & -1 & 0 \\
0 & 0 & 0 & 1%
\end{array}%
\right) .
\end{equation}

A straightforward calculation shows that, the eigenequation $H|\Psi
\rangle =E|\Psi \rangle $ of the Hamiltonian $H$ can be re-written as
\begin{equation}
\sum_{l^{\prime }}\mathbf{H}_{ll^{\prime }}^{(0)}\beta _{l^{\prime
}}+\sum_{l^{\prime }}\mathbf{V}_{ll^{\prime }}\beta _{l^{\prime }}=E\beta
_{l}.  \label{she}
\end{equation}%
Here for each given value of $\left( l,l^{\prime }\right) $, $\mathbf{H}%
_{l,l^{\prime }}^{(0)}$ and $\mathbf{V}_{l,l^{\prime }}$ are $4$-dimensional
matrixes and defined as%
\begin{equation}
\mathbf{H}_{l,l^{\prime }}^{(0)}=\mathcal{A}\delta _{l,l^{\prime }}+\left(
\frac{\mathcal{B}}{2}+\frac{i\mathcal{C}}{2}\right) \delta _{l-1,l^{\prime
}}+\left( \frac{\mathcal{B}}{2}-\frac{i\mathcal{C}}{2}\right) \delta
_{l+1,l^{\prime }}  \label{h0}
\end{equation}%
and
\begin{equation}
\mathbf{V}_{l,l^{\prime }}=\left(
\begin{array}{cccc}
0 & 0 & 0 & 0 \\
0 & 0 & 0 & -\sqrt{2}g \\
0 & 0 & 0 & 0 \\
0 & -\sqrt{2}g & 0 & 0%
\end{array}%
\right) \delta _{l,0}\delta _{l^{\prime },0},  \label{v}
\end{equation}%
with the matrixes $\mathcal{A}$, $\mathcal{B}$ and $\mathcal{C}$ given by%
\begin{eqnarray}
\mathcal{A} &=&\left(
\begin{array}{cccc}
0 & \sqrt{2}g & 0 & 0 \\
\sqrt{2}g & \Delta & 0 & \sqrt{2}g \\
0 & 0 & \Delta & 0 \\
0 & \sqrt{2}g & 0 & 2\Delta%
\end{array}%
\right) , \\
\mathcal{B} &=&2\xi \cos \left( K/2\right) \left(
\begin{array}{cccc}
2 & 0 & 0 & 0 \\
0 & 1 & 0 & 0 \\
0 & 0 & 1 & 0 \\
0 & 0 & 0 & 0%
\end{array}%
\right) , \\
\mathcal{C} &=&-2\xi \sin \left( K/2\right) \left(
\begin{array}{cccc}
0 & 0 & 0 & 0 \\
0 & 0 & 1 & 0 \\
0 & 1 & 0 & 0 \\
0 & 0 & 0 & 0%
\end{array}%
\right) .
\end{eqnarray}

The above discussion shows that the Schr\"{o}dinger equation for the
relative motion of two polaritons in our system is equivalent to Eq.~(\ref%
{she}) and the boundary condition (\ref{c1}) for the coefficient $\beta _{l}$%
. It is obvious that Eq.~(\ref{she}) has a
same form as the stationary Schr\"{o}dinger equation for a scattering
problem. Then we can understand the matrix $\mathbf{H}_{l,l^{\prime }}^{(0)}$
as the \textquotedblleft free-Hamiltonian" of the two-polariton relative
motion, and $\mathbf{V}_{l,l^{\prime }}$ as the \textquotedblleft
interaction" between these two polaritons. Therefore, we can use the
standard technique of quantum scattering problem to solve Eq.~(\ref{she}),
and then find the scattering states and bound states of the two polaritons
in our system.

Now we comment on the physical picture given by the inter-polariton
interaction $\mathbf{V}_{l,l^{\prime }}$. It is clear that, the symbol $%
l\left( l^{\prime }\right) $ is an abbreviation of $m-n$ in Eq. (\ref{psi}),
and thus describes the relative position of two polaritons. Since $\mathbf{V}%
_{l,l^{\prime }}$ takes non-zero value only when $l=l^{\prime }=0$, it is a
two-polariton contact potential. Namely, two polaritons interact with each
other when they are in the same cavity. To understand the physical meaning
of this contact potential, we consider a extreme case where the photon
cannot tunnel between different cavities. In that case, the total
Hamiltonian becomes $H=\sum_{j=-\infty }^{\infty }H_{JC}^{(j)}$ with the
JC Hamiltonian $H_{JC}^{(j)}=\Delta \sigma
_{ee}^{(j)}+g\sigma _{eg}^{\left( j\right) }a_{j}+h.c.$. It is clear that $%
H_{JC}^{(n)}$ can be diagonalized in the subspace with $\sigma
_{ee}^{(j)}+a_{j}^{\dagger }a_{j}=$ $n$ $(n=0,1,2...)$, and the relevant
eigen-energies are $\varepsilon _{JC}^{(\pm )}(n)=(\Delta \pm \sqrt{\Delta
^{2}+4g^{2}n})/2$. If two polaritons appear in two different cavities, the
total energy of the two polaritons can take the ($i,j$)-independent
value $\varepsilon
_{JC}^{(\pm )}(1)+\varepsilon _{JC}^{(\pm )}(1)$. However, if the two polaritons appear in the same cavity, the energy of the two polaritons take the value $\varepsilon_{JC}^{(\pm)}(2) \neq \varepsilon_{JC}^{(\pm)}(1)+\varepsilon_{JC}^{(\pm)}(1)$. Therefore, due to the
nonlinearity of the spectrum of JC Hamiltonian, the energy of two
polaritons changes when they are in the same cavity. That is the origin of
the effective contact interaction $\mathbf{V}_{l,l^{\prime }}$ of two
polaritons in our system.

\section{Scattering and bound states of two polaritons}

In the above section, we find that the two-polariton problem in the 1D
cavity array is described by Eq.~(\ref{she}) with boundary condition (\ref{c1}%
). In this section we show our approach for solving Eq.~(\ref{she}) and
derive the scattering states and bound states of two polaritons.

\subsection{Scattering states}

Now we calculate the two-polariton scattering state. To this end, we first
analytically solve the eigenequation%
\begin{equation}
\sum_{l^{\prime }}\mathbf{H}_{l,l^{\prime }}^{(0)}\beta _{l^{\prime
}}^{(0)}=E\beta _{l}^{(0)}  \label{he1}
\end{equation}%
of $\mathbf{H}_{l,l^{\prime }}^{(0)}$ and find the \textquotedblleft
free-motion" state of the two polaritons. Due to the translation symmetry of
$\mathbf{H}_{l,l^{\prime }}^{(0)}$, a basic solution of Eq.~(\ref{he1})
takes the form $e^{iql}F\left( q\right) $ with $q\in \left[ -\pi ,\pi
\right) $ the relative momentum of the two polaritons and the $l$-independent
vector $F\left( q\right) $ satisfies%
\begin{equation}
\left[ \mathcal{A}+\mathcal{B}\cos q+\mathcal{C}\sin q\right] F\left(
q\right) =EF\left( q\right) .  \label{epsi}
\end{equation}%
It is obvious that, for a given value of $q$, Eq.~(\ref{epsi}) has four
solutions for $F\left( q\right) $. We denote these solutions as $F_{\alpha}\left(
q\right) $ with $\alpha $ taking the values $AA,AB,BA,BB$.
For a physical meaning of these solutions,
see the final paragraph in this section.
The analytical
expression of $F_{\alpha }(q)$ is given in Appendix A. Straightforward
calculation also shows that, the eigen-energy $E(\alpha ,q)$ with respect to
$e^{iql}F_{\alpha }\left( q\right) $ can be expressed as
\begin{equation}
E\left( \alpha ,q\right) =\mathcal{E}_{u}\left( \Delta _{0},\delta
_{1}\right) +\mathcal{E}_{v}\left( \Delta _{0},\delta _{2}\right) .
\label{a1}
\end{equation}%
Here the symbols $u,v$ can take the values $A,B$, and related with $\alpha $
via the relationship $\alpha =uv$. In Eq. (\ref{a1}) we also have $\Delta
_{0}=2\xi \cos q\cos (K/2)$ and $\delta _{1,2}=\Delta \mp 2\xi \sin q\sin
(K/2)$.The function $\mathcal{E}_{A,B}(x,y)$ is defined as
\begin{equation}
\mathcal{E}_{A,B}\left( x,y\right) =\frac{1}{2}\left( x+y\right) \pm \frac{1%
}{2}\sqrt{\left( x-y\right) ^{2}+4g^{2}}.
\end{equation}%
We further define the vector $\beta _{l}^{(0)}\left( \alpha ,q\right) $ as%
\begin{equation}
\beta _{l}^{(0)}\left( \alpha ,q\right) =\frac{1}{2}\left[ e^{iql}F_{\alpha
}\left( q\right) +e^{-iql}\mathcal{T}F_{\alpha }\left( q\right) \right] .
\end{equation}%
It is easy to prove that $\beta _{l}^{(0)}\left( \alpha ,q\right) $
satisfies both Eq.~(\ref{he1}) and the boundary condition (\ref{c1}), and
then can be considered as the \textquotedblleft free-motion" state of the
two polaritons.

Now we consider the scattering wave function $\beta _{l}^{(+)}\left( \alpha
,q\right) $ with respect to the incident wave function $\beta
_{l}^{(0)}\left( \alpha ,q\right) $. $\beta _{l}^{(+)}$ is given by the
Lippmman-Schwinger equation%
\begin{equation}
\beta _{l}^{(+)}\left( \alpha ,q\right) =\beta _{l}^{(0)}\left( \alpha
,q\right) +\sum_{l_{1},l_{2}}\mathbf{G}_{l,l_{1}}^{(0)}\mathbf{V}%
_{l_{1},l_{2}}\beta _{l_{2}}^{(+)}\left( \alpha ,q\right) ,  \label{lse}
\end{equation}%
where the Green's function $\mathbf{G}_{l^{\prime \prime },l^{\prime
}}^{(0)} $ is the solution of the equation%
\begin{equation}
\sum_{l^{\prime \prime }}\left[ E(\alpha ,q)+i0^{+}-\mathbf{H}_{l,l^{\prime
\prime }}^{(0)}\right] \mathbf{G}_{l^{\prime \prime },l^{\prime
}}^{(0)}=I\delta _{l,l^{\prime }}
\end{equation}%
with $I$ the $4$-dimensional identical matrix. The straightforward
calculations (see, e.g., chapter 9 of Ref.~\cite{scbook}) with the
Lippmman-Schwinger equation (\ref{lse}) show that $\beta _{l}^{(+)}\left(
\alpha ,q\right) $ takes the form%
\begin{eqnarray}
\beta _{l}^{(+)}\left( \alpha ,q\right) & =&\beta _{l}^{(0)}\left( \alpha
,q\right) +\sum_{\gamma }f(\gamma \leftarrow \alpha ,q)e^{i\lambda _{\gamma
}l}F_{\gamma }(\lambda _{\gamma })  \notag \\
&&\ \ \ \ \ \ \ \ \ \ \ \ \ \ \ \ \ \ \ \ \ \ \ \ \ \ \ \ \ \ \ \ \ \ \
(l>0),  \label{st1} \\
\beta _{l}^{(+)}\left( \alpha ,q\right) & =&\mathcal{T}\beta
_{-l}^{(+)}\left( \alpha ,q\right)\ \ \ \ \ \ \ \ \ \ \ \ \ \ \ \ \ \ \
(l<0), \\
\beta _{0}^{(+)}\left( \alpha ,q\right) &=&\left( s_{p},s_{+},0,0\right)
^{T}.  \label{st4}
\end{eqnarray}%
Here $F_{\gamma }(\lambda _{\gamma })$ is given by Eq.~(\ref{fab}) and the
parameters $(\gamma ,\lambda _{\gamma })$ satisfy $E(\gamma ,\lambda
_{\gamma })=E(\alpha ,q)$. Namely, $\lambda _{\gamma }$ can be obtained via
the equation%
\begin{equation}
\det \left\vert \mathcal{A}+\mathcal{B}\cos \lambda _{\gamma }+\mathcal{C}%
\sin \lambda _{\gamma }-IE(\alpha ,q)\right\vert =0.  \label{ee}
\end{equation}%
In addition, $\lambda _{\gamma }$ also satisfies the conditions%
\begin{eqnarray}
\mathrm{Im}\lambda _{\gamma } &\geq &0;  \label{cc1} \\
\frac{\partial }{\partial \lambda _{\gamma }}E(\gamma ,\lambda _{\gamma })
&>&0\text{ \ when }\mathrm{Im}\lambda _{\gamma }=0.  \label{cc2}
\end{eqnarray}%
It is easy to prove that Eqs. (\ref{ee}-\ref{cc2}) have three solutions.
Thus, the summation in Eq. (\ref{st1}) includes three terms. In Eq.~(\ref{st1}), the
factor $f(\alpha \leftarrow \alpha ,q)$ is the elastic scattering coefficient, while $%
f(\gamma \leftarrow \alpha ,q)$ for the term with $\gamma \neq \alpha $ and $%
\mathrm{Im}\lambda _{\gamma }>0$ is the inelastic scattering coefficient which
describes the inter-channel transition induced by the scattering process.

For a given incident wave function $\beta _{l}^{(0)}\left( \alpha ,q\right) $%
, we can obtain the scattering wave function $\beta _{l}^{(+)}\left( \alpha
,q\right) $ with the following two steps. First, solve Eqs. (\ref{ee}-\ref%
{cc2}) and find the three solutions for $(\gamma ,\lambda _{\gamma })$.
Second, substitute expressions (\ref{st1}-\ref{st4}) into equations $%
\sum_{l^{\prime }}(\mathbf{H}_{l,l^{\prime }}^{(0)}+\mathbf{V}_{l,l^{\prime
}})\beta _{l^{\prime }}^{(+)}=E\beta _{l}^{(+)}$ with $l=0,1$, and obtain
the values of $s_{+}$, $s_{p}$ and the values of the three coefficients $%
f(\gamma \leftarrow \alpha ,q).$

In the end of this subsection, we discuss the physical meaning of the
two-polariton scattering state. To this end, we first consider a state $%
|\Phi \rangle =A_{K/2+q}^{\dag }A_{K/2-q}^{\dag }|G\rangle $ with the
operator $A_{k}^{\dagger }$ defined in Eq.~(\ref{ab}). The physical meaning
of the state $|\Phi \rangle $ is that there are two excitonic
polaritons of type $A$ with momentums $K/2+q$ and $K/2-q$. It is apparent
that the state $|\Phi \rangle $ can be written in the form of Eq.~(\ref{psi}%
). We denote the $\beta $-coefficient or the relative wave function of $|\Phi \rangle $ as $\beta _{l}(\Phi )$. A straightforward
calculation shows that we have $\beta _{l}(\Phi )=\beta _{l}^{(0)}\left(
AA,q\right) $. Furthermore, it can also be proved that $\varepsilon
_{A,K/2+q}+\varepsilon _{A,K/2-q}=E\left( AA,q\right) $. Namely, the total
energy of the two excitonic polaritons in the state $|\Phi \rangle $ is the
same as the energy of the incident state with wave function $\beta _{l}^{(0)}\left(
AA,q\right) $. Thus, the incident wave function
$\beta _{l}^{(0)}\left( AA,q\right) $ can be
considered as the relative wave function of the ``free motion" of two
excitonic polaritons of type $A$ with total momentum $K$ and relative
momentum $q$. Similar discussions can also be done for the states with two
excitonic polaritons of the other types. Therefore, $\beta _{l}^{(0)}\left(
uv,q\right) $ and $\beta _{l}^{(+)}\left( uv,q\right) $ with $u,v=A,B$ can
be considered as the relative wave function of the free motion and
scattering state of two excitonic polaritons of types $u$ and $v$ with total
momentum $K$ and relative momentum $q$.

\subsection{Bound states}

Now we consider the bound states of two polaritons. The energy $E_{b}$ and
the wave function $\beta _{l}^{(b)}$ of the bound state are determined by
the eigenequation%
\begin{equation}
\sum_{l^{\prime }}\mathbf{H}_{l,l^{\prime }}^{(0)}\beta _{l^{\prime
}}^{(b)}+\sum_{l^{\prime }}\mathbf{V}_{l,l^{\prime }}\beta _{l^{\prime
}}^{(b)}=E_{b}\beta _{l}^{(b)}  \label{eb}
\end{equation}%
and the boundary condition%
\begin{equation}
\lim_{|l|\rightarrow \infty }\beta _{l}^{(b)}=0.  \label{bb}
\end{equation}%
Similar as in the above subsection, the solution of Eqs. (\ref{eb}, \ref{bb}%
) takes the form%
\begin{eqnarray}
\beta _{l}^{(b)} &=&\sum_{j=1}^{3}b_{j}e^{i\lambda _{j}l}F_{\alpha
_{j}}(\lambda _{j}),\ \ \ \ \ \ \ \ \ (l>0)  \label{bf1} \\
\beta _{l}^{(b)} &=&\mathcal{T}\beta _{-l}^{(b)},\ \ \ \ \ \ \ \ \ \ \ \ \ \
\ \ \ \ \ \ \ \ \ \ \ (l<0)  \label{bf2} \\
\beta _{0}^{(b)}&=&\left( b_{p},b_{+},0,0\right) ^{T},  \label{bf3}
\end{eqnarray}%
where $F_{\alpha }$ is defined in our above subsection and $(\alpha
_{j},\lambda _{j})$ is given by the relationship $E(\alpha _{j},\lambda
_{j})=E_{b}$ and satisfies the equations
\begin{eqnarray}
\det \left\vert \mathcal{A}+\mathcal{B}\cos \lambda _{j}+\mathcal{C}\sin
\lambda _{j}-IE_{b}\right\vert &=&0;  \label{d2} \\
\mathrm{Im}\lambda _{j} &>&0.
\end{eqnarray}%
In addition, substituting Eqs. (\ref{bf1}-\ref{bf3}) into equations Eq.~(\ref%
{eb}) with $l=0,1$, one can get the homogeneous linear equations for the
coefficients $b_{1,2,3}$, $b_{p}$ and $b_{+}$. When there is a two-polariton
bound state, the determinant of the coefficient matrix of these equations
should be zero. From this condition, we can obtain the energy $E_{b}$ of the
bound state. Substituting the value of $E_{b}$ into these linear equations
and using the normalization condition $\sum_{l}\beta _{l}^{(b)\dagger }\beta
_{l}^{(b)}=1$, we can obtain the coefficients $b_{1,2,3}$, $b_{p}$
and $b_{+}$.

\begin{figure}[tbh]
\centering
\subfigure{
    \includegraphics[bb=51 270 385 565, width=6.5cm, clip]{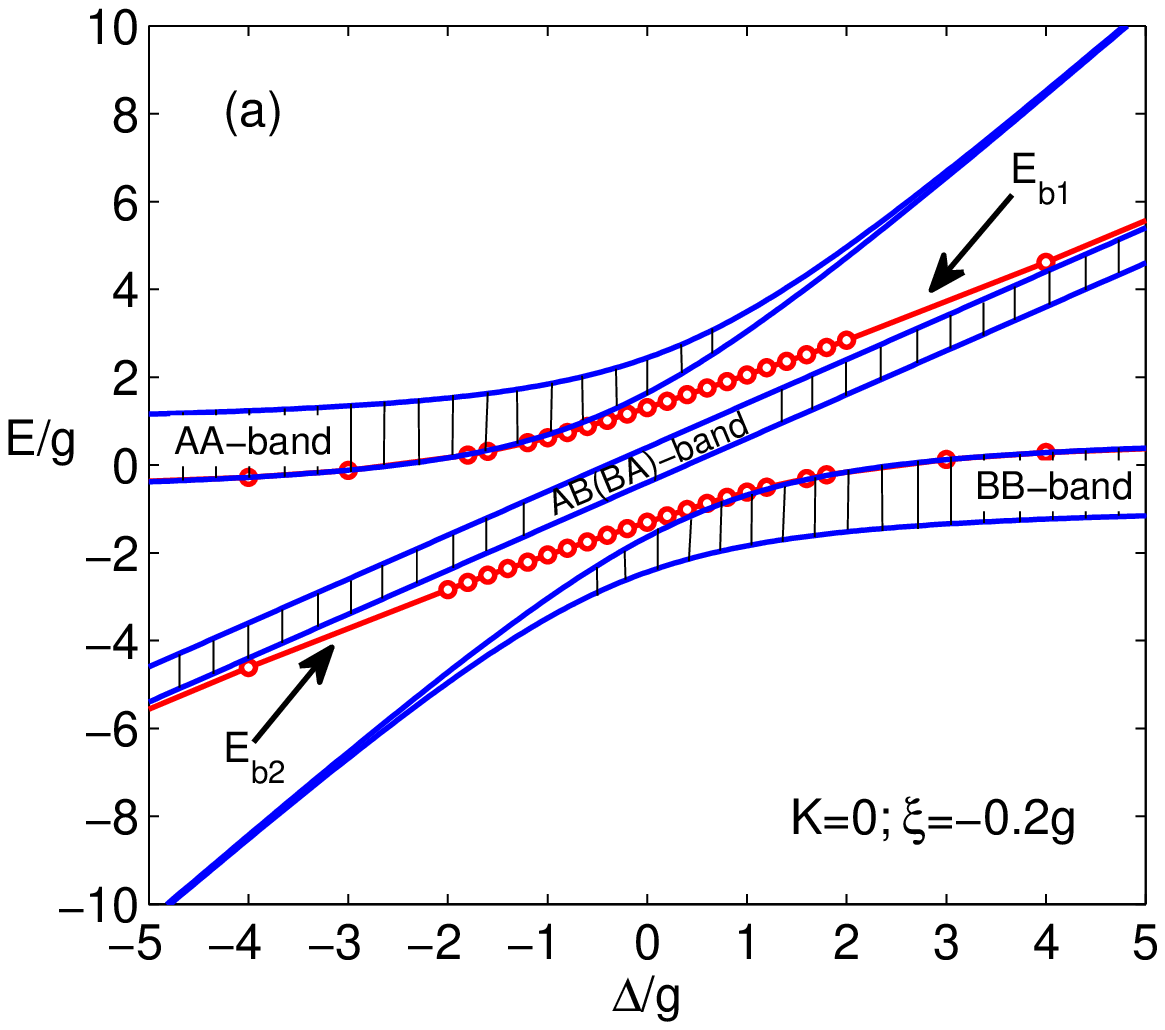}}
\subfigure{
    \includegraphics[bb=51 270 385 565, width=6.5cm, clip]{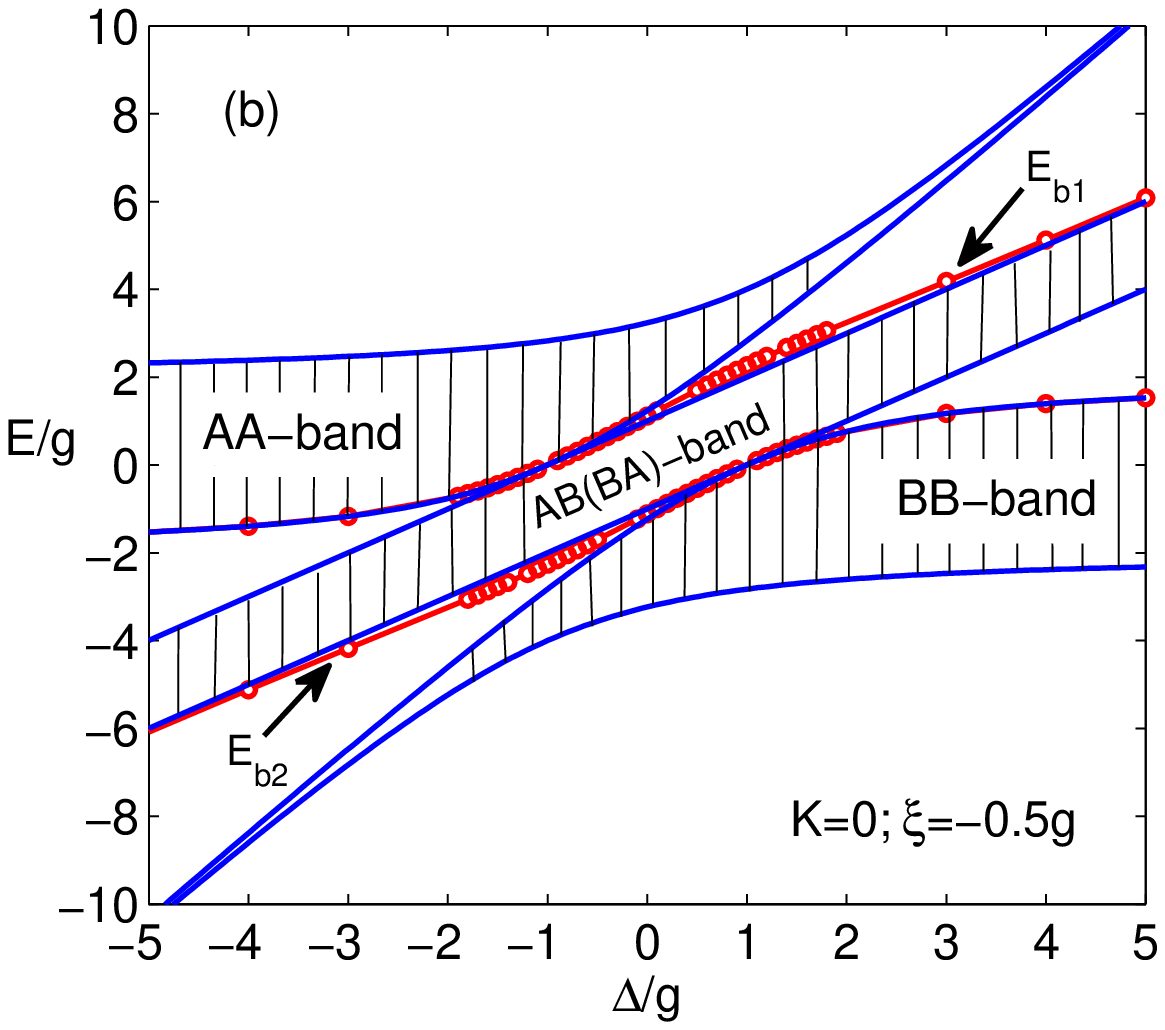}}
\subfigure{
    \includegraphics[bb=51 270 385 565, width=6.5cm, clip]{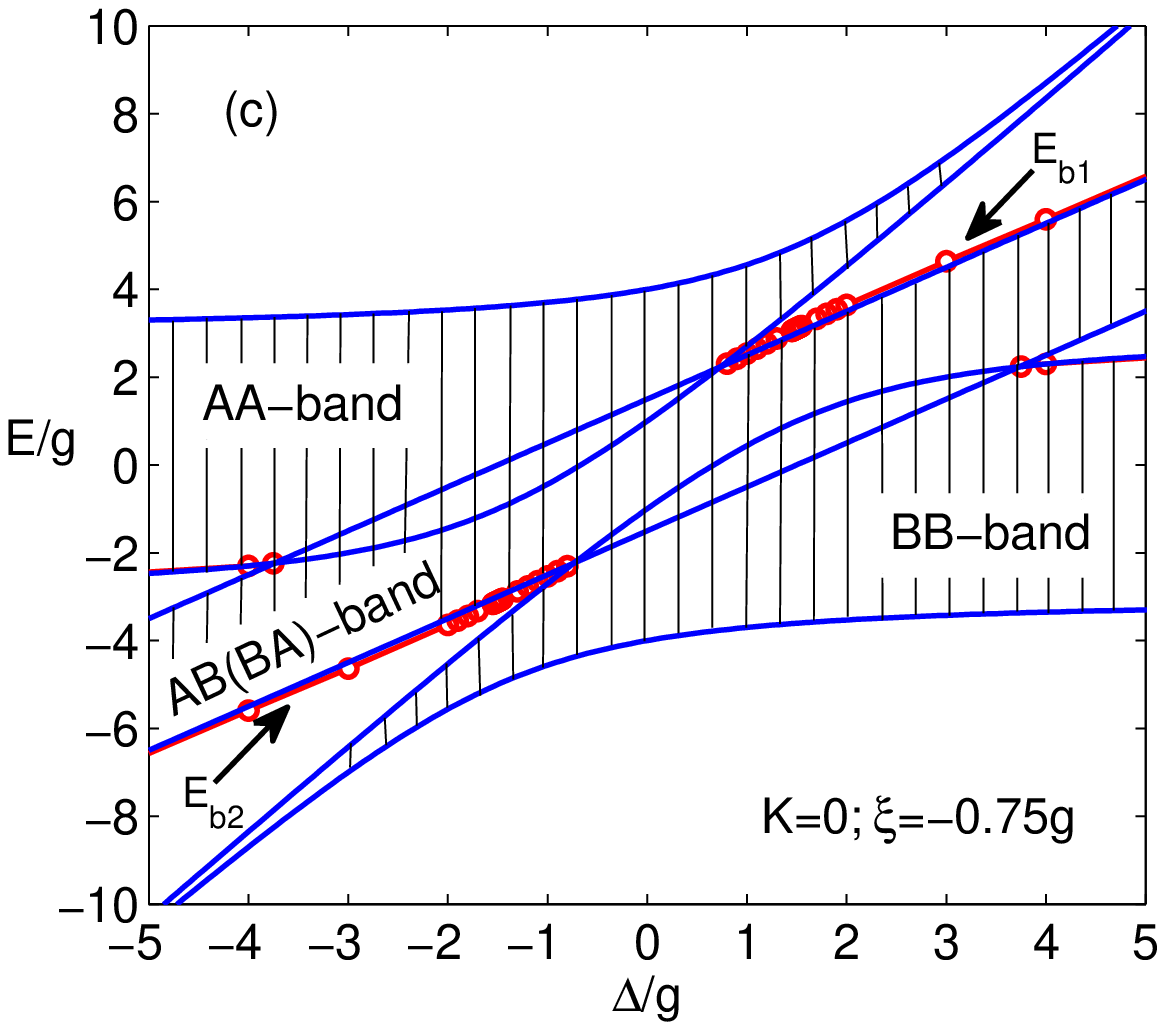}}
\caption{(Color online) The energy bands of scattering states (regions with
shadow) and eigen-energies of bound states (red circles) as functions of the
detuning $\Delta $. The figures are plotted for the cases with total
momentum $K=0$ and photonic hopping intensity $\protect\xi=-0.2g $ (a), $%
-0.5g $ (b) and $=-0.75g $ (c), with $g$ the photon-TLS coupling intensity.}
\label{fig4}
\end{figure}

\begin{figure}[tbh]
\centering
\subfigure{
    \includegraphics[bb=36 430 291 639, width=6.3cm, clip]{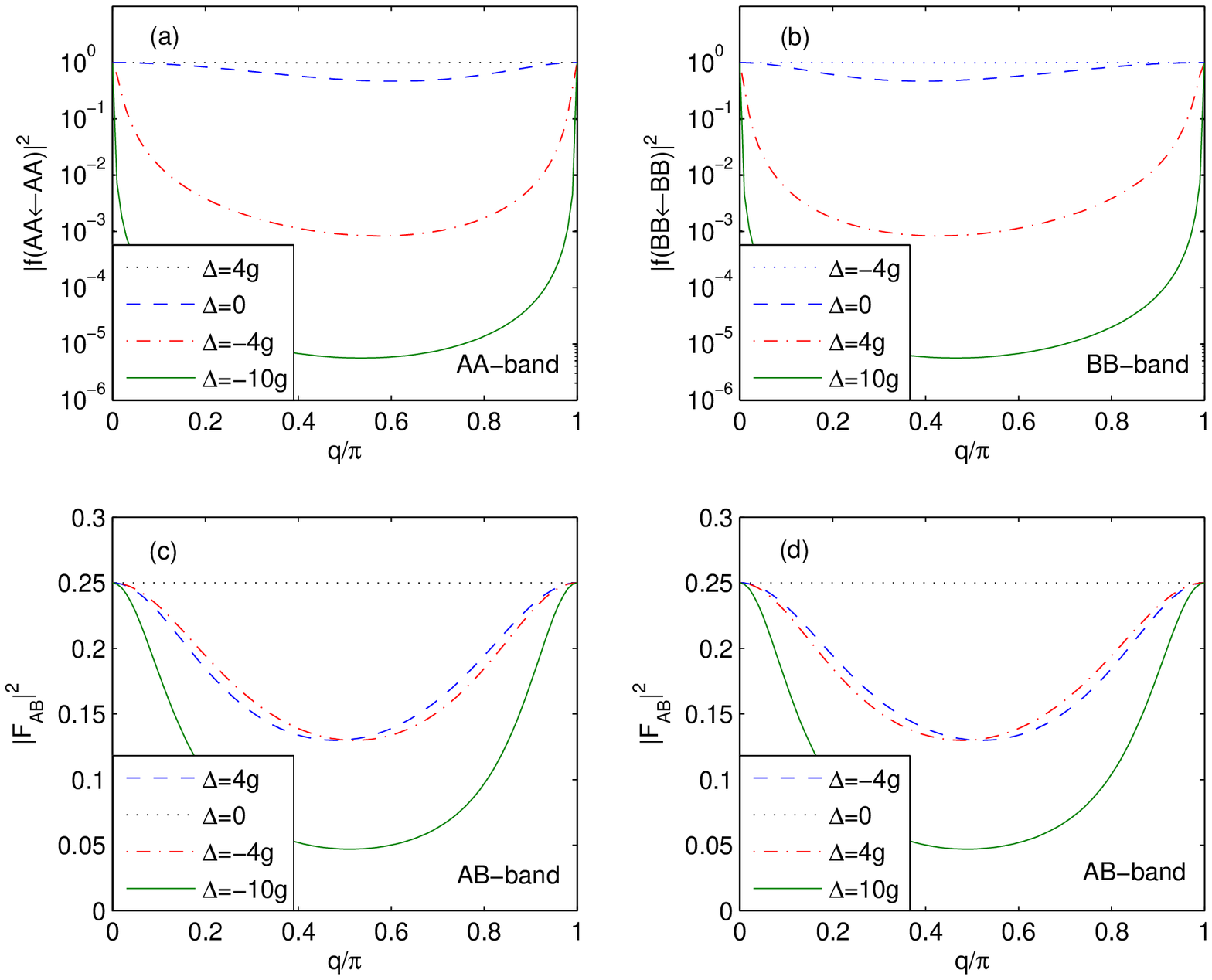}}
\subfigure{
    \includegraphics[bb=316 430 567 639, width=6.3cm, clip]{scat_amp.eps}}
\subfigure{
    \includegraphics[bb=36 206 291 415, width=6.3cm, clip]{scat_amp.eps}}
\caption{(Color online) Norm-square of two-polariton scattering coefficient as
a function of relative momentum $q$ of the incident state with total
momentum $K=0$ and photonic hopping intensity $\protect\xi =-0.2g$. (a)
Norm-square of the elastic scattering coefficient $f(AA\leftarrow AA,q)$. (b)
Norm-square of the elastic scattering coefficient $f(BB\leftarrow BB,q)$. (c)
Norm-square of the elastic and inelastic scattering coefficient $%
f(AB\leftarrow AB,q)$ and $f(BA\leftarrow AB,q)$. In our cases we have $%
f(AB\leftarrow AB,q)=f(BA\leftarrow AB,q)$, and then denote the value of the
two amplitudes as $F_{AB}$ in the figures.}
\label{fig4}
\end{figure}

\section{two-polariton problem with zero center-of-mass momentum}

In the above section, we show our analytical approach for the calculations
of two-polariton scattering states and bound states in a 1D cavity array.
With our method, one can derive the scattering coefficient and bound states of
two polaritons in the systems with any detuning $\Delta $ and coupling
parameters $(g,\xi )$. Now we show the results of our calculations. For
simplicity, in this paper we only consider case with the total momentum $K=0
$.

As discussed above, the energy of the two-polariton scattering state $\beta
_{l}^{(+)}\left( uv,q\right) $ with $u,v=A,B$ is $E(uv,q)$ defined Eq.~(\ref%
{a1}) . In the following, we call the region of value of $E(uv,q)$ with $%
q\in[-\pi,\pi)$ as ``$uv$-band". Then the energies of the scattering states
are located in $AA$-, $AB$-, $BA$-, and $BB$-bands. Furthermore, we have $%
E(AB,q)=E(BA,-q)$. Thus the $AB$-band and $BA$-band totally overlap with
each other, and we have three energy bands for the two-polariton scattering
states. In Fig.~2, these three bands are shown as the regions with shadow.

We further find that, in our system there are two bound states with wave
functions $\beta _{l}^{(b1)}$ and $\beta _{l}^{(b2)}$ and energies $E_{b1}$
and $E_{b2}$ which are located in the gap between $AA$- and $AB$-band, and
the one between $AB$- and $BB$-band, respectively. In Fig.~2 we demonstrate
the energies $E_{b1} $ and $E_{b2}$ as functions of the detuning $\Delta $.

In the large-detuning cases with $|\Delta |>>|g|,|\xi |$, the effects of
photon-TLS coupling become weak. As a result, the creation and
annihilation of photons, as well as the quantum transitions between the
ground and excited states of TLS, can be adiabatically eliminated. In
addition, the effective coupling between photons in different cavities, as
well as the one between the excited TLSs in different cavities, can appear
in our system. Then one of the two bound states can be approximated as the
two-photon bound state, and another one becomes the bound state of two
excitations of TLS. Furthermore, the energies of the two bound states become
very close to the borders of the energy bands of the scattering states.

The above analysis are verified by our quantitative calculation. Our results
show that, when $\Delta <0$ and $|\Delta |>>|g|,|\xi |$, the energy $E_{b1}$
approaches to the lower limit of the $AA$-band (Fig.~2), and we have $\beta
_{l}^{(b1)}\approx (p_{l},0,0,0)^{T}$. Therefore, the two-polariton bound
state $|\Psi _{b1}\rangle $ satisfies
$|\Psi _{b1}\rangle \approx \sum_{m,l}p_{l}a_{m}^{\dagger }a_{m+l}^{\dagger
}|G\rangle $, and it is approximately a two-photon
bound state. On the other hand, when $\Delta >>|g|,|\xi |$, the energy $%
E_{b1}$ approaches to the upper limit of the $AB$-band (Fig.~2), and we have
$\beta _{l}^{(b1)}\approx (0,0,0,a_{l})^{T}$. Therefore, the two-polariton
bound state $|\Psi _{b1}\rangle $
satisfies $|\Psi _{b1}\rangle \approx \sum_{m,l}t_{l}\sigma
_{eg}^{(m)}\sigma _{eg}^{(m+l)}|G\rangle $, and becomes approximately a
bound state of two excitations of TLS.

Similar analysis can be done for the bound state $|\Psi _{b2}\rangle $ with
wave function $\beta _{l}^{(b2)}$ and energy $E_{b2}$. As shown in Fig.~2,
the energy $E_{b2}$ approaches to the upper limit of $BB$-band and the lower
limit of $AB$-band in the limits $\Delta \rightarrow \pm \infty $,
respectively. Furthermore, $|\Psi _{b2}\rangle $ is approximately a
two-photon bound state when $\Delta >>|g|,|\xi |$, and approximately a bound
state of two excitations of TLS when $\Delta <0$ and $|\Delta |>>|g|,|\xi |$.

In the 1D low-energy scattering problem of two non-relativistic particles in
the continuous space (e.g., the scattering problem of a non-relativistic particle
 in a delta potential $g\delta(x)$ with $g<0$), it is well-known that, when the
  energy of the bound state is close to the lower limit of the scattering energies,
the scattering coefficient  (reflection coefficient) becomes very small.
In our system, we find
similar resonance results. As shown above, in our cases the band of the
scattering energies has both the lower border and the upper border. We find
that when the bound-state energy is close to either of these two borders,
the relevant scattering coefficient $f$ becomes very small.

For the elastic scattering coefficient $%
f(AA\leftarrow AA,q)$ of the scattering state in the $AA$ band, as
shown in Fig.~3(a), when $\Delta <0$ and $|\Delta |>>|g|,|\xi |$ and $E_{b1}$
is close to the lower bound of $AA$-band, we have $f(AA\leftarrow
AA,q)\approx 0$. On the other hand, when $\Delta >>|g|,|\xi |$ and $E_{b1}$
is relatively far from the lower bound of $AA$-band we get the result $%
|f(AA\leftarrow AA,q)|\approx 1$. Our results show that, the effective
interaction between two excitonic polaritons of kind $A$ is negligible in
the limit $\Delta \rightarrow -\infty $, and becomes strongly repulsive when
$\Delta \rightarrow \infty $.

For the elastic scattering coefficients $%
f(BB\leftarrow BB,q)$, $f(AB\leftarrow AB,q)$ and the inelastic scattering
coefficient $f(BA\leftarrow AB,q)$, as shown in Fig.~3(b), we have $%
f(BB\leftarrow BB,q)\approx 0$ when $\Delta >>|g|,|\xi |$ and $E_{b2}$ is
close to the upper bound of $BB$-band. Finally, for the scattering states in
the $AB(BA)$ band, according to our calculation (Fig.~2),  the energy $E_{b1}
$ is close to the upper bound of the $AB$-band when $\Delta >>|g|,|\xi |$,
while $E_{b2}$ is close to the lower bound of the $AB$-band when $\Delta <0$
and $|\Delta |>>|g|,|\xi |$. The corresponding resonance phenomenon is
illustrated in Fig.~3(c),  where it is shown that in both of these two
regions we always have $f(AB\leftarrow AB,q)=f(AB\leftarrow BA,q)\approx 0$.

\section{Conclusion and discussion}

In this paper we derive the effective interaction between two polaritons in
a 1D cavity-array coupled to TLSs, and provide an analytical method for the
calculation of scattering and bound states in such a system. We find that in a cavity array there is an effective contact
interaction between two polaritons, which is induced by the nonlinearity of the JC Hamiltonian.  The interaction is totally determined by the photon-TLS coupling strength $g$, and independent of the photonic hopping
intensity $\xi$.
For two polaritons with zero center-of-mass momentum, we find that there are
two bound states in the gaps between the energy bands of the polaritons, and
the 1D resonance phenomenon can appear when the photon-TLS detuning is large
enough. Our result is helpful for the research of
the few-body and many-body physics of
polaritons in cavity arraies. In particular,
 for the dilute polariton gas where the average polariton number in each cavity is much smaller than one,
the many-body physics is dominated by the two-polariton contact interaction.

\begin{acknowledgments}
This work was supported by National Natural Science Foundation of China
under Grants No. 11074305, 11222430, NKBRSF of China under Grants No.
2012CB922104, and the Research Funds of Renmin University of China
(10XNL016).
\end{acknowledgments}

\appendix
\addcontentsline{toc}{section}{Appendices}\markboth{APPENDICES}{}
\begin{subappendices}

\section{The expression of $F_{\alpha}\left( q\right)$}

In this appendix we provide the analytical expression of
the vector $F_{\alpha}\left( q\right)$ ($\alpha=AA,\ AB,\ BA,\ BB$) defined in Sec. III. With straightforward calculation, we solve Eq.~(\ref{epsi}) analytically and
get the expression of $F_{uv}\left( q\right)$ ($u,v=A,B$):
\begin{equation}
F_{uv}\left( q\right) =\left(
\begin{array}{c}
C_{gu }^{(1)}C_{gv }^{(2)} \\
\frac{1}{\sqrt{2}}\left( C_{eu }^{\left( 1\right) }C_{gv}^{\left(
2\right) }+C_{gu }^{\left( 1\right) }C_{ev}^{\left( 2\right)
}\right) \\
\frac{1}{\sqrt{2}}\left( C_{eu}^{\left( 1\right) }C_{gv}^{\left(
2\right) }-C_{gu}^{\left( 1\right) }C_{ev}^{\left( 2\right)
}\right) \\
C_{eu}^{\left( 1\right) }C_{ev}^{\left( 2\right) }%
\end{array}%
\right) ,  \label{fab}
\end{equation}%
where the coefficients $C_{e(g),A(B)}^{\left( 1,2\right) }$ are defined as $%
C_{e(g),A(B)}^{\left( 1,2\right) }=f_{e(g),A(B)}\left( \delta _{1,2}-\Delta
_{0}\right) $ with $\delta _{1,2}$ and $\Delta_0$ defined in Sec. IIIA. Here the functions $f_{e(g),A(B)}\left( x\right) $ are defined as
\begin{eqnarray}
f_{g,A(B)}\left( x\right) &=&\frac{2g}{\sqrt{4g^{2}+\left\vert x\pm \sqrt{%
x^{2}+4g^{2}}\right\vert ^{2}}}; \\
f_{e,A(B)}\left( x\right) &=&\frac{x\pm \sqrt{x^{2}+4g^{2}}}{\sqrt{%
4g^{2}+\left\vert x\pm \sqrt{x^{2}+4g^{2}}\right\vert ^{2}}}.
\end{eqnarray}

\end{subappendices}

\end{document}